\documentclass[aps,prb,twocolumn,amsmath,amssymb,floatfix,superscriptaddress]{revtex4-2}
\usepackage{blindtext}
\usepackage{epsfig}
\usepackage{color}
\usepackage{amsmath}
\usepackage{amsfonts}
\usepackage{amssymb}
\usepackage{graphicx}%
\usepackage{url}
\usepackage[breaklinks]{hyperref}
\usepackage[english]{babel}
\usepackage{esvect}
\usepackage[normalem]{ulem}
\usepackage[inline]{enumitem}
\usepackage{xcolor}
\usepackage{xfrac}
\usepackage{dsfont}
\usepackage{bm}
\usepackage{array}
\newcolumntype{C}[1]{>{\centering\let\newline\\\arraybackslash\hspace{0pt}}m{#1}}
\usepackage{footnote}
\usepackage{isomath}
\DeclareMathAlphabet\mathbfcal{OMS}{cmsy}{b}{n}

\hypersetup{colorlinks,citecolor=blue,filecolor=blue,linkcolor=blue,urlcolor=blue}

\bibpunct{[}{]}{,}{n}{}{}

\begin{document}

\title{Topological Signatures of Magnetic Phase Transitions with Majorana Fermions through Local Observables and Quantum Information}

\author{Karyn Le Hur}
\affiliation{CPHT, CNRS, Institut Polytechnique de Paris, Route de Saclay, 91128 Palaiseau, France}
\author{Fan Yang}
\affiliation{Department of Physics, Stockholm University, AlbaNova University Center, 10691 Stockholm, Sweden}
\affiliation{CPHT, CNRS, Institut Polytechnique de Paris, Route de Saclay, 91128 Palaiseau, France}
\author{Magali Korolev}
\affiliation{CPHT, CNRS, Institut Polytechnique de Paris, Route de Saclay, 91128 Palaiseau, France}

\begin{abstract}
The one-dimensional (1D) $J_1-J_2$ quantum spin model can be viewed as a strong-coupling analogue of the Schrieffer-Su-Heeger model with two inequivalent alternating Ising couplings along the wire, associated to the physics of resonating valence bonds. 
Similar to the quantum Ising model, which differently presents a long-range N\' eel ordered phase, this model also maps onto a p-wave superconducting wire which shows a topological phase transition with the emergence of low-energy Majorana fermions. 
We show how signatures of the topological phase transition for the p-wave superconducting wire, i.e. a half Skyrmion, are revealed through local (short-range) spin observables and their derivatives related to the capacitance of the pairing fermion model. 
Then, we present an ``edge'' correspondence through the edge spin susceptibility in the $J_1-J_2$ model revealing that the topological phase transition is a metal of Majorana fermions. We justify that the spin magnetization at an edge at very small transverse magnetic field is a good marker of the topological invariant and of Majorana zero modes. We identify a correspondence between the quantum information of resonating valence bonds and the charge fluctuations in a p-wave superconductor through our method ``the bipartite fluctuations''. Physical properties of this 1D model are in fact robust when including additional interactions, which is optimistic for practical applications e.g. in quantum circuits.
\end{abstract}
\maketitle

Quantum phase transitions \cite{Sachdev,Carr} are attracting attention since the last decades. The 1D quantum Ising model \cite{DeGennes,Pfeuty} is a paradigm which can be solved through an analogy with the topological p-wave superconducting wire \cite{Kitaev}. This system is e.g. realized in quantum circuits associated to topologically protected Majorana edge modes \cite{XMi}. The phase transition is usually signalled through the occurrence of long-range (two-point) correlation functions going to zero with a critical exponent $\beta=\frac{1}{8}$ 
for the spin magnetization \cite{Pfeuty}. The superconducting p-wave Kitaev wire is rather described through a topological invariant jumping from one to zero between the topological phase with Majorana edge modes and the trivial phase corresponding to strong on-site pairing for the Majorana fermions \cite{Kitaev}. Then, we may question the relation between spin observables and topological properties. The present work is also motivated from the fact that the $J_1-J_2$ spin chain \cite{Feng,KarynArianeFan} or equivalently the 1D compass model \cite{Sun}, corresponding to alternating couplings along $x$ and $y$ directions on successive bonds, 
is also an interesting 1D model for the quest of $\mathbb{Z}_2$ two-dimensional (2D) quantum spin liquids \cite{Kitaevhoneycomb} and resonating valence bonds \cite{LiangDoucotAnderson}, presenting a correspondence onto the same p-wave superconducting wire \cite{FanKirillKaryn}. 
In this Letter, we introduce {\it local} spin observables, e.g. short-range spin correlation functions to reveal the topological properties of the p-wave superconducting wire, with an emphasis on the $J_1-J_2$ model which remains to be fully understood. Understanding this model is also important as a relevant platform for quantum information through Majorana fermions related to the Kitaev honeycomb spin model \cite{Kitaevhoneycomb} and to quantum spin models with fractional topology including the topological model of two spheres \cite{EphraimBrianKaryn,HH,KarynReview}. We will introduce a correspondence between the topological invariant and the poles of the Bloch sphere \cite{KarynReview} associated to the Bardeen-Cooper-Schrieffer (BCS) model in momentum space \cite{FrederickLoicKaryn,FrederickLoicOlesiaKaryn}. Derivatives of these short-range correlators will mark the location of the phase transition related to the capacitance of the p-wave superconductor. We propose a probe at the edge for the $J_1-J_2$ model, the local spin susceptibility when applying a small (transverse) field along $z$ axis;  this will evidence that at the topological quantum phase transition the edge magnetic susceptibility shows a logarithmic singularity reminiscent of the two-channel Kondo model (2CKM) \cite{NozieresBlandin,EmeryKivelson,SenguptaGeorges,GiamarchiShraimanClarke}, revealing low-energy Majorana fermions. 
The edge magnetization encodes the jump of the topological invariant.

The description of quantum phase transitions in magnetic materials in terms of quantum information probes is also attracting attention related to the long-range quantumness of the correlations e.g. through the quantum Fisher information \cite{Zoller,Tennant,Si}. We clarify a correspondence between the structure of the resonating valence bonds in the $J_1-J_2$ model through bipartite charge fluctuations \cite{LoicKarynChristophe} that we introduced as a marker of the entangled nature of many-body systems \cite{FluctuationsReview}. 

The 1D $J_1-J_2$ model takes the form \cite{Feng,KarynArianeFan}
\begin{equation}
{\cal H}=\sum_{j=2m-1} (J_1\sigma_j^x \sigma_{j+1}^x + J_2 \sigma_{j+1}^y \sigma_{j+2}^y),
\end{equation}
with $m\geq 1$. We assume $J_1,J_2>0$ and a even number of sites to have symmetric edges. It will be useful to introduce hearafter two Majorana fermions per site $c_j$ and $d_j$ through the Jordan-Wigner transformation \cite{Feng,KarynArianeFan} such that the Majorana fermion $d_j$ will be {\it free} producing a $\frac{1}{2}\ln 2$ thermodynamical entropy per site. Measuring such an entropy associated to a Majorana fermion is attracting attention e.g. with quantum information probes \cite{2CKMmeasure} and in materials through the 2CKM \cite{2CKMmaterials}. Per bond, the system then presents a $\ln 2$ residual entropy. Suppose that $J_1>0$; then the system presents two equiprobable ground states on a $J_1$ link $\uparrow_x\downarrow_x$ and $\downarrow_x\uparrow_x$ such that this indeed produces a $\ln 2$ entropy. Similarly, if $J_2>0$ then the system presents two equiprobable ground states $\uparrow_y\downarrow_y$ and $\downarrow_y\uparrow_y$ for an adjacent link, which also produces a $\ln 2$ residual entropy. The two phases for $J_1>J_2$ and $J_2>J_1$ can then be viewed as quantum spin liquids with resonating valence bonds forming preferably along $x$ or $y$ direction. The Jordan-Wigner transformation precisely takes the form $\sigma_j^+ = a^{\dagger}_j\prod_{i<j} (-\sigma_j^z)$, $\sigma_j^- = a_j \prod_{i<j} (-\sigma_j^z)$ with $\sigma_j^z=2 a^{\dagger}_j a_j -1$. It is then judicious to introduce the Majorana fermions with alternating definitions on sites $j=2m-1$ and sites $j=2m$ such that for $j=2m-1$, we have \cite{FanKirillKaryn} $c_j=i(a^{\dagger}_j-a_j)$, $d_j=a_j^{\dagger}+a_j$ and for $j=2m$, we 
have $c_j=a_j^{\dagger}+a_j$, $d_j=i(a^{\dagger}_j-a_j)$ and $c_j^2=d_j^2=1$ such that $\{a_j,a_j^{\dagger}\}=1$. The Hamiltonian then reads \cite{FanKirillKaryn}
\begin{equation}
\label{model}
{\cal H}=\sum_{j=2m-1} (-i)(J_1 c_j c_{j+1} -J_2 c_{j+1} c_{j+2}).
\end{equation}

To acquire an understanding of the nature of the phase transition in real space --- that was not mentioned in previous works e.g. in Refs. \cite{Feng,KarynArianeFan,FanKirillKaryn} --- we observe that the Hamiltonian for $J_1=J_2=J$ is equivalent to
\begin{equation}
\label{Hc}
{\cal H}=iJ\sum_{j=2m-1} c_{j+1}(c_j+c_{j+2}).
\end{equation}
If we take the view of the long-distance physics $c_{j+2}=c_j+2a\partial_j c_j$ with $a$ the lattice spacing. If we label the Majorana fermions $\gamma_A$ on odd sublattice and $\gamma_B$ on even sublattice --- to account for the fact that even and odd sites are inequivalent  e.g. the definition of $c_j$ is different --- the Hamiltonian reads ${\cal H} = 2iJa\int dx \gamma_B(x)\partial_x \gamma_A(x)$ modulo a constant term in energy. Then, we recognize the form of quantum field theory for the p-wave superconducting wire at the phase transition \cite{FrederickLoicKaryn}. Then, ${\cal H}$ can be diagonalized through the Majorana modes $\sqrt{2}\gamma_{L,R}=(\gamma_B\pm \gamma_A)$ such that ${\cal H} =iJa\int dx (\gamma_L\partial_x \gamma_L - \gamma_R \partial_x \gamma_R)$. This gives rise to a $\frac{1}{2}$ degree of freedom (left or right) in the specific heat \cite{Affleck} in relation to the 2D classical Ising model \cite{ItzyksonDrouffe}, which refers to a {\it metal of Majorana fermions} \cite{Karyn1999}. The band structure shows a linear dispersion and reveals zero-energy Majorana fermions. This $\frac{1}{2}$ justifies the central charge $c=\frac{1}{2}$ in the entanglement
 entropy \cite{FanKirillKaryn}. 
 
 Even though we address a specific model of interactions on each link (bond), properties of the two phases and of the phase transition presented below will be in fact very robust to perturbations
 associated e.g. to interactions along $z$ direction; see Supplementary Material \cite{SM} and also results from the Density Matrix Renormalization Group (DMRG) approach \cite{DatasLink}. Hereafter, we present the results for the model (\ref{model}) which then encodes a larger variety of realizable Hamiltonians (see \cite{SM}). The proofs for the generalized models remain very similar \cite{SM}.

\begin{figure}[t]
\includegraphics[width=8.5cm]{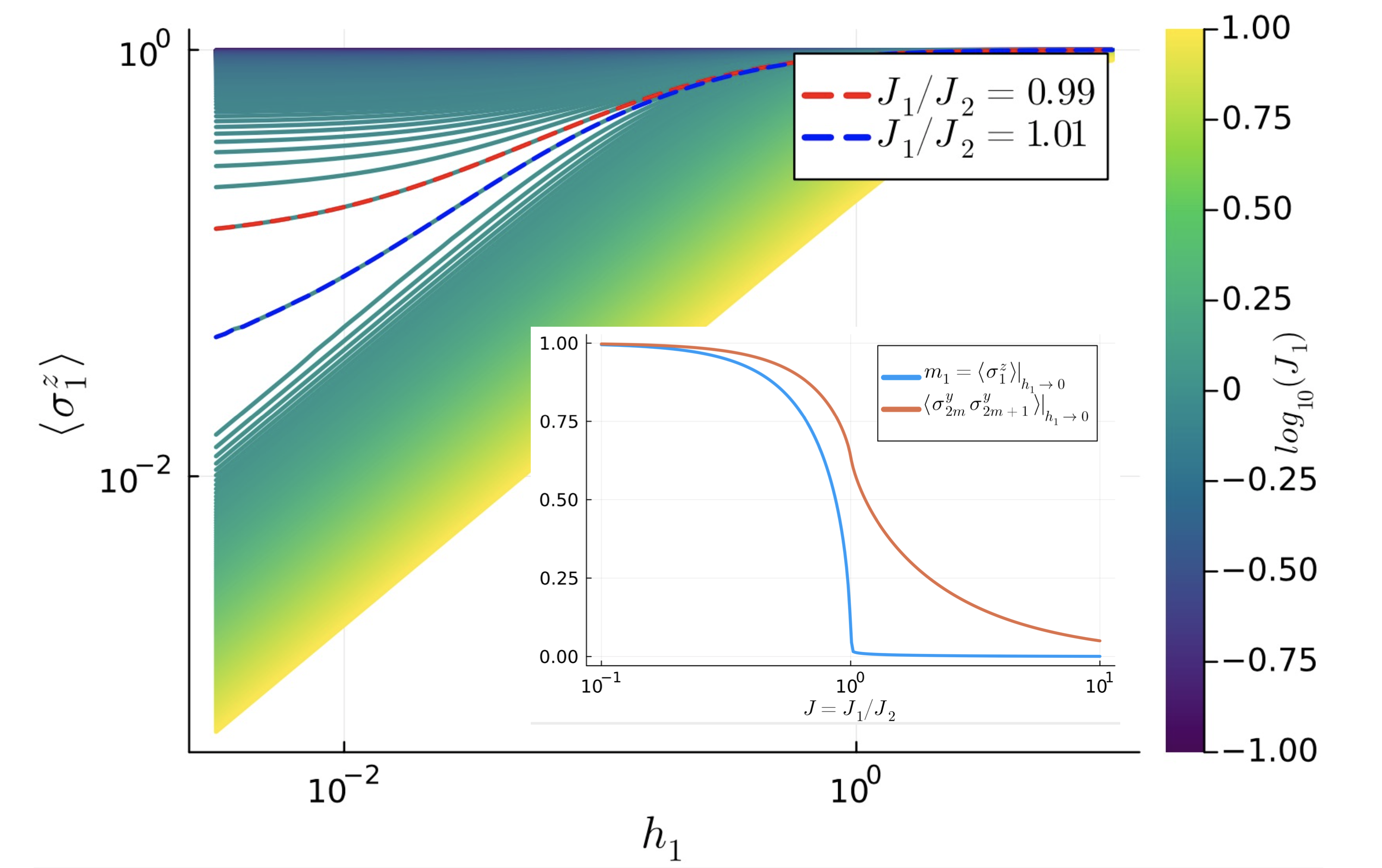}
\caption{Spin magnetization at the edge obtained with DMRG (400 sites) \cite{DatasLink} as a function of $h_1$ and $J_1$ with $J_2=1$. Comparison between the edge magnetization in blue when $h_1\rightarrow 0$ and $\langle \sigma_{2m}^y \sigma_{2m+1}^y\rangle$ correlation
function in orange in Fig. \ref{Clarifynumbers} (with $\langle \sigma_{2m}^y \sigma_{2m+1}^y\rangle\rightarrow - \langle \sigma_{2m}^y \sigma_{2m+1}^y\rangle$).}
\label{magnetization}
\end{figure}

The two phases for $J_1<J_2$ and $J_1>J_2$ are different in terms of the {\it local spin magnetization at an edge}, which then represents a local order parameter for the structure of Majorana zero modes and of the topological phase transition. Suppose we apply a magnetic field a site $1$ along $z$ axis. This results in the perturbation $\delta H=+i h_1 c_1 d_1=-h_1 \sigma_1^z$. When $J_1\rightarrow 0$, at $h_1=0$, the two Majorana fermions $c_1$ and $d_1$ are free at zero energy. As soon as $h_1=0^+$, the ground state is characterized with a fixed value of the parity operator $i c_1 d_1=-1$ such that the spin magnetization shows a plateau at unit magnetization. We will show below in Eq. (\ref{topomagic}) that within the topological phase for $J_1<J_2$, $\langle i c_2 c_3\rangle = \langle\sigma_{2}^y \sigma_3^y\rangle \rightarrow -{\cal C}$ with ${\cal C}$ being the {\it quantized} topological number. This implies that $\langle i c_1 c_2\rangle=0$, as a topological protection of the edge properties. In this way, the response at the edge to the local magnetic field evidences the topological phase transition, as seen in Fig. \Ref{magnetization} through the DMRG approach \cite{DatasLink}. The edge magnetization precisely jumps when $J_1=J_2$ through a diverging logarithmic singularity that we will elucidate thoroughly hereafter in Eq. (\ref{chi1}). When including the magnetic field $h_1$ when $J_1\gg J_2$, we can write the magnetic terms associated to $c_1$ as $-i c_1(-h_1 d_1+ J_1 c_2) = -i c_1 \eta \sqrt{h_1^2 +J_1^2}$ with the composite Majorana fermion $\eta = \frac{-h_1 d_1+ J_1 c_2}{\sqrt{h_1^2+J_1^2}}$. Minimization of energy here corresponds to $\langle  i c_1 \eta \rangle \rightarrow 1$. At small $h_1$, $\langle \sigma_1^z\rangle = \langle -i c_1 d_1\rangle = \frac{h_1}{2J_1}$ whereas $i c_1 c_2\rightarrow +1$ (to linear order in $h_1$); see yellow line in Fig. \ref{magnetization}. 

Below, we analyze the topological bulk-edge correspondence. It is not easy to find spin correlation functions which are non-zero beyond nearest neighbors because $\langle c_i d_i\rangle=0$, $\langle i d_i d_{j}\rangle=0$ for $i\neq j$ such that $\langle \sigma_i^z \sigma_{i+1}^z\rangle=0$. It is natural to introduce the {\it nearest-neighbor correlators} 
\begin{eqnarray}
\label{marker}
\sigma_{2m-1}^x \sigma_{2m}^x = -i c_{2m-1} c_{2m} \\ \nonumber
\sigma_{2m}^y \sigma_{2m+1}^y = i c_{2m} c_{2m+1}.
\end{eqnarray}
To evaluate the mean values of these spin observables, we introduce the {\it bond fermions} \cite{FanKirillKaryn}
\begin{equation}
\psi_m = \frac{1}{2}(c_{2m-1} + i c_{2m}),
\end{equation}
with a doubled lattice spacing $l=2a$. This leads to
\begin{eqnarray}
{\cal H} = \sum_{m=1}^M -2J_1(\psi^{\dagger}_m \psi_m -\frac{1}{2}) &-& J_2(\psi^{\dagger}_m\psi_{m+1}+h.c.) \\ \nonumber
&-& J_2(\psi^{\dagger}_m \psi_{m+1}^{\dagger} +h.c.).
\end{eqnarray}
The number of sites $M$ for the $\psi$ fermions is equal to the number of links $J_1$ or $J_2$.
This quantum spin model from the bond fermions $\psi$ is then equivalent to a (lattice) p-wave Kitaev superconducting Hamiltonian \cite{Kitaev} with hopping term $t=J_2$, pairing term $\Delta=-J_2$ and chemical potential $\mu=2J_1$. 
The topological phase in the spin language corresponds to $J_1<J_2$ (i.e. to $\mu<2t$) associated to the free Majorana fermion $c_1$ and the trivial phase refers to $J_1>J_2$ (i.e. to $\mu>2t$ with $\mu>0$).
The correspondence with the charge density of fermions reads
\begin{eqnarray}
\label{charge}
\sigma_{2m-1}^x \sigma_{2m}^x = -\psi^{\dagger}_m\psi_m +\psi_m \psi^{\dagger}_m. \\ \nonumber
\end{eqnarray}
Associated to the {\it kinetic term} and to the {\it pairing term} 
\begin{eqnarray}
\label{spinobservables}
\langle \sigma_{2m}^y \sigma_{2m+1}^y\rangle &=& -D_{kin} + D_{sc} \nonumber \\
\langle \sigma_{2m-1}^x \sigma_{2m}^z \sigma_{2m+1}^z \sigma_{2m+2}^x\rangle &=& -D_{kin}-D_{sc}
\end{eqnarray}
where $D_{kin}=\langle \psi^{\dagger}_m\psi_{m+1}+h.c. \rangle = -\frac{i}{2}\langle c_{2m}c_{2m+1} - c_{2m-1}c_{2m+2}\rangle$ and $D_{sc} =\langle \psi_m \psi_{m+1}+h.c.\rangle=\frac{i}{2}\langle c_{2m} c_{2m+1} + c_{2m-1}c_{2m+2}\rangle$. Deep in the topological phase at $J_1\rightarrow 0$ and deep in the trivial phase at $J_2\rightarrow 0$, we must have $\langle \sigma_{2m-1}^x \sigma_{2m}^z \sigma_{2m+1}^z\sigma_{2m+2}^x\rangle =0$ due to the structure of the Hamiltonian in Eq. (\ref{model}) whereas $\langle \sigma_{2m}^y \sigma_{2m+1}^y\rangle$ should vary from $-1$ to $0$ when increasing the ratio $J_1/J_2$. 
We precisely question a possible relation between these observables and the topological properties linked to the edge spin magnetization. 

\begin{figure}[t]
\includegraphics[width=8cm]{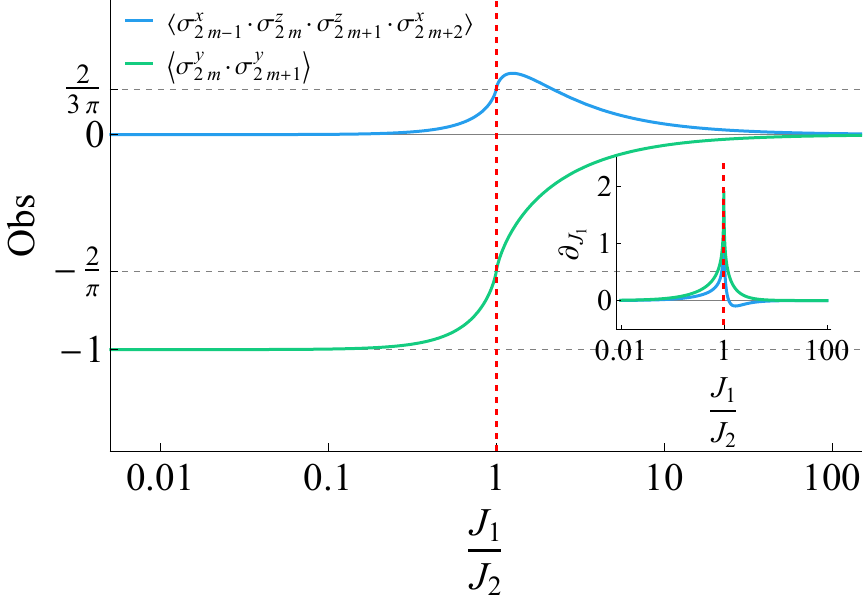}
\caption{Observables from Eqs. (\ref{spinobservables}), (\ref{observables}) and DMRG \cite{DatasLink}.}
\label{Clarifynumbers}
\end{figure}

From the BCS wavefunction that we remind in the Supplementary Material \cite{SM}, it is then possible to derive general analytical formulas for $D_{kin}$ and $D_{sc}$ in terms of geometrical properties of a spin-1/2 particle \cite{FrederickLoicOlesiaKaryn,SM}
\begin{eqnarray}
\label{observables}
D_{kin} &=& \frac{1}{M} \sum_k \cos(kl) \langle S^z\rangle =  \frac{1}{M} \sum_k \cos(kl) \cos \theta_k \\ \nonumber
D_{sc} &=& \frac{1}{M} \sum_k \sin(kl) \langle S^y\rangle = -\frac{1}{M}\sum_k \sin(kl)\sin \theta_k.
\end{eqnarray}
The Anderson pseudo-spin variable \cite{Anderson,Aoki} is associated to the $2\times 2$ matrix Hamiltonian in momentum space and allows for a correspondence with the Bloch sphere i.e. $S^z=\psi^{\dagger}_k \psi_k - \psi_{-k}\psi_{-k}^{\dagger}$, $S^x=\psi^{\dagger}_k \psi^{\dagger}_{-k} +\psi_{-k}\psi_k$ and $S^y=-i(\psi^{\dagger}_k \psi^{\dagger}_{-k} - \psi_{-k} \psi_k)$ \cite{KarynReview}. Here, $\theta_k$ is the polar angle on the sphere associated to the spin-1/2 \cite{SM}
\begin{eqnarray}
\label{poles}
\cos\theta_k &=& \frac{J_1+J_2\cos kl}{E(k)} \\ \nonumber
\sin\theta_k e^{-i\varphi} &=& -ie^{i\phi}\frac{J_2\sin(k l)}{E(k)}
\end{eqnarray}
with 
\begin{equation}
E(k)=\sqrt{J_1^2+J_2^2+2J_1 J_2\cos(kl)},
\end{equation}
associated to the quasiparticle energy spectrum. Here, $\theta$ can be mapped onto a function of the wave-vector $k$ whereas the azimuthal angle $\varphi$ on the sphere corresponds to the superfluid phase associated to the order parameter $\Delta$ \cite{KarynReview,FrederickLoicKaryn}. We introduce a phase $\phi+\pi$ to $\Delta$ such that $\psi^{\dagger}_k \psi^{\dagger}_{-k}\rightarrow -e^{i \phi}\psi^{\dagger}_k \psi^{\dagger}_{-k}$ and we set $e^{-i\varphi}=-i e^{ i\phi}$. The additional $\pi$ phase is to compensate for the fact that $\Delta=-J_2$. We present the solutions of Eqs. (\ref{observables}) and of spin observables in Eqs. (\ref{spinobservables}) in Fig. \ref{Clarifynumbers}, which agree with the DMRG Approach \cite{DatasLink}. Now, we aim to clarify the topological character of these spin observables.

The global topological invariant within the topological phase is precisely described through the poles of the sphere from the Anderson pseudo-spin component $\langle S^z\rangle=\langle\psi^{\dagger}_k \psi_k - \psi_{-k} \psi^{\dagger}_{-k}\rangle=\cos\theta_k$ 
\cite{KarynReview,FrederickLoicKaryn}, such that
\begin{eqnarray}
\label{invariant}
{\cal C} &=& \frac{1}{2}(\langle S^z(kl=0)\rangle - \langle S^z(kl=\pi)\rangle) \\ \nonumber
&=& \frac{1}{2}(sgn(J_1+J_2) -sgn(J_1-J_2)).
\end{eqnarray}
In this formula, $\langle S^z\rangle$ should be thought of as the function $\cos\theta_k$ in Eq. (\ref{poles}) that depends on the variable $kl$. 
This topological invariant clearly shows a jump from one when $J_1=0$ to zero when $J_2=0$ at $J_1=J_2$. This invariant associated to a spin-1/2 and to the Nambu space of the p-wave superconducting wire then classifies the phases and quantum phase transition. 
It is closely related to the spin magnetization at the edge in Fig. \ref{magnetization}.

For $J_1=0$ and $J_2>0$ the polar angle satisfies $\theta_k=kl\in [0;\pi]$ and the model describes the presence of a Skyrmion or magnetic monopole with an integer invariant
\begin{eqnarray}
\label{sgntopo}
{\cal C} &=& \frac{1}{2}(\langle S^z(kl=0)\rangle - \langle S^z(kl=\pi)\rangle) \nonumber \\
&=& sgn(J_2)=2D_{kin} = -2D_{sc}.
\end{eqnarray}
As can be seen in Fig. \ref{Clarifynumbers}, the spin correlation function along $y$ axis shows a nice plateau such that this characterizes the integer topological marker (invariant) 
\begin{equation}
\label{topomagic}
\langle \sigma_{2m}^y \sigma_{2m+1}^y\rangle =-D_{kin}+D_{sc} = -sgn(J_2)=-{\cal C},
\end{equation}
protecting the $c_1$ and $d_1$ edge modes  in relation to $\langle \sigma_1^z\rangle$.

At the quantum phase transition $J_1=J_2=J$, $\theta_k=\frac{kl}{2}\in[0;\frac{\pi}{2}]$ from Eq. (\ref{poles}). The quantum phase transition can be understood as a topological invariant on a half sphere 
equivalent to a half Skyrmion \cite{KarynReview}
\begin{eqnarray}
\label{topohalf}
{\cal C}_{1/2}=sgn(J)\frac{1}{2}.
\end{eqnarray}
At the quantum phase transition of the $J_1-J_2$ model, we find that local kinetic and pairing correlators precisely reveal this half-topological invariant in Eq. (\ref{topohalf}):
\begin{eqnarray}
\label{fourtwo}
\langle \sigma_{2m}^y \sigma_{2m+1}^y\rangle = -D_{kin}+D_{sc} = -\frac{4{\cal C}_{1/2}}{\pi} \hskip 0.5cm  \\ \nonumber
\langle \sigma_{2m-1}^x \sigma_{2m}^z \sigma_{2m+1}^z \sigma_{2m+2}^x\rangle = -D_{kin}-D_{sc} = \frac{4{\cal C}_{1/2}}{3\pi}. 
\end{eqnarray}
These values agree with the direct numerical integration in Fig. \ref{Clarifynumbers} and with DMRG results. At this stage, we can also clarify a useful correspondence with the quantum Ising model.
From Eqs. (\ref{observables}), we verify that 
\begin{equation}
\label{formula}
\langle \sigma_{2m}^y \sigma_{2m+1}^y\rangle = -\frac{l}{\pi}\int_0^{\frac{\pi}{l}} \frac{J_2+J_1 \cos(kl)}{E(k)} dk
\end{equation}
which is related to the spin magnetization integral associated to the transverse field in the quantum Ising model \cite{OsborneNielsen}. This can be understood from a direct mapping between the $J_1-J_2$ chain and the transverse Ising model on the dual lattice with doubled lattice spacing $2a=l$ \cite{Feng} such that $\sigma_j^x = \tau_{j-1}^x\tau_j^x$ and $\sigma_j^y = \Pi_{k=j}^{2M} \tau_k^y$. The spin magnetization $\langle \tau_{2m}^y\rangle$ then is equivalent to $\langle \sigma_{2m}^y \sigma_{2m+1}^y\rangle$. We find that $\langle \sigma_{2m-1}^x \sigma_{2m}^x\rangle$ shows a symmetric behavior compared to $\langle \sigma_{2m}^y \sigma_{2m+1}^y\rangle$ when inverting the roles of $J_1$ and $J_2$ in Fig. \ref{Clarifynumbers}. 

To illustrate the usefulness of the geometrical approach on the sphere, we mention that when $J_1=J_2=J$ 
\begin{equation}
\langle \sigma_{2m}^y \sigma_{2m+1}^y\rangle = -sgn(J)\frac{2}{\pi}\int_0^{\frac{\pi}{2}} \cos \theta d\theta.
\end{equation}
Through a simple change of variables $\theta'=\frac{\pi}{2}-\theta$ we obtain a half Skyrmion via the integral of the Berry curvature $F_{\theta\varphi}=\frac{\sin\theta}{2}$ for a spin-1/2 particle on a half sphere \cite{KarynReview}
\begin{equation}
\label{19}
\langle \sigma_{2m}^y \sigma_{2m+1}^y\rangle = -sgn(J)\frac{4}{\pi}\int_0^{\frac{\pi}{2}} F_{\theta\varphi} d\theta = -\frac{4}{\pi}{\cal C}_{1/2}.
\end{equation}

\begin{figure}[t]
\includegraphics[width=7cm]{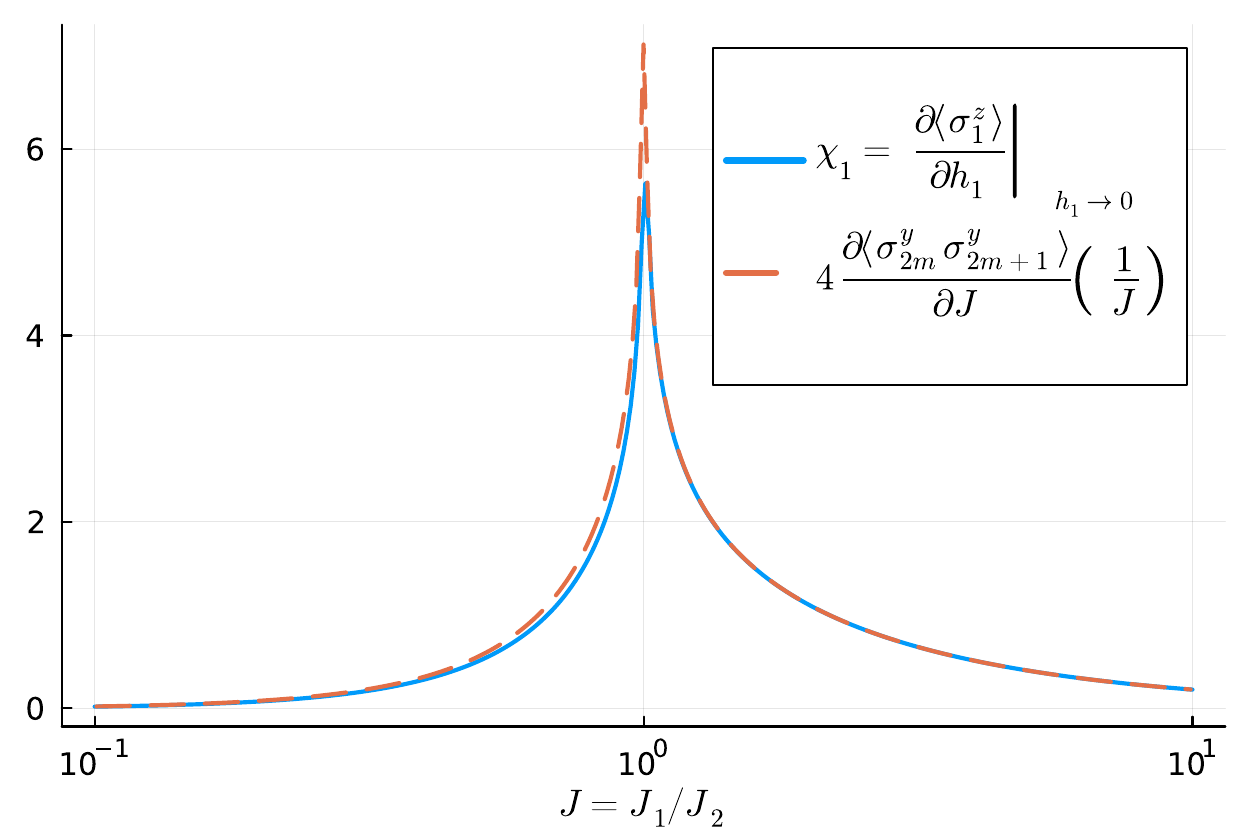}
\caption{{\it Bulk-edge correspondence}: Edge spin susceptibility with $h=h_1$ from DMRG which diverges logarithmically when $J_1=J_2$ 
(see also Eq. (\ref{chi1})). Logarithmic divergence of the derivative associated to the short-range spin correlation function(s) which also agrees with Eq. (\ref{transition}).}
\label{susceptibility}
\end{figure}

It is then useful to draw a parallel between the occurrence of Majorana fermions $\gamma_L$ and $\gamma_R$ introduced after Eq. (\ref{Hc}) and the correspondence on the sphere when $J_1=J_2$.
Since the gap for the $\psi$ fermions is closing at $kl=\pi$ at the quantum phase transition then this is equivalent to zero-energy  Majorana fermions in the equatorial plane at $\theta=\frac{\pi}{2}$. 
The quasiparticle modes at energy $\pm J_2\sin(kl)$, which goes to zero linearly close to $kl=\pi$, are $\eta_{kl=\pi} = \frac{1}{\sqrt{2}}(\psi_{\pi} \pm i \psi^{\dagger}_{\pi})$. The addition of these two modes represents $\psi^{\dagger}_{\pi}$.
The wave-vectors at $kl=\pi$ and $kl=-\pi$ are identical from applications of the translation lattice vector. Introducing the Majorana fermions $\gamma_A$ and $\gamma_B$ such that $\psi_m=\frac{1}{2}(\gamma_{A,2m-1}+i\gamma_{B,2m})$ then $\eta_k$ also describes the dynamics of the $\gamma_L$ and $\gamma_R$ Majorana fermions at $kl=\pi$ with $\eta_{\pi}=\frac{1}{2\sqrt{2}}(1\pm i)(\gamma_{A,\pi}\pm\gamma_{B,\pi})$ as for the transition in the p-wave superconductor \cite{FrederickLoicKaryn}.

Then, we report that the derivative of this observable in Eq. (\ref{formula}) shows a dominant logarithmic singularity  
\begin{equation}
\label{transition}
\frac{1}{2\pi} \int_0^{\pi} \frac{1}{\cos\frac{u}{2}}du \sim \frac{1}{\pi}\int_0^{\frac{\pi}{2}} \frac{du'}{u'},
\end{equation}
in Figs. \ref{Clarifynumbers} and \ref{susceptibility}
which is also useful to locate the transition point in Figs. \Ref{Clarifynumbers} and \Ref{susceptibility}. For $J_1>J_2$, the topological invariant is ${\cal C}=0$ and $D_{kin}$ and $D_{sc}$ go slowly to zero in Fig. \ref{Clarifynumbers}. For the p-wave superconductor, it is in fact possible to re-sum in a series all the pairing and kinetic correlators in space to re-build the half topological invariant at the transition \cite{FrederickLoicOlesiaKaryn}. In the $J_1-J_2$ model, two-point spin correlations are short-range i.e. between nearest neighbors. 
It is yet possible to re-build all correlators in the p-wave superconductor in terms of spin operators accompanied with Jordan-Wigner strings. Then, we observe the same {\it power-law} decay for the valence bond correlation function \cite{FanKirillKaryn} when $J_1=J_2$ and for the pairing and kinetic 
correlation functions that will motivate us to also introduce Eq. (\ref{iQ}).  
The key point here is that the logarithmically diverging derivative of the short-range spin correlations in the $J_1-J_2$ model also detects the transition.  Since $J_1$ represents the chemical potential of the $\psi$ fermions, the derivative of these short-range spin correlations encode the behavior of the capacitance of the p-wave superconducting wire (for any $m$) through Eq. (\ref{charge}). When refermionizing the quantum Ising model ${\cal H}=\sum_{i=2m} J_2 \tau_i^y + J_1\tau_{i-2}^x \tau_{i}^x$ in a usual way \cite{Sachdev,Pfeuty}, then $\tau_{2m}^y=\tilde{\sigma}_i^z=2a^{\dagger}_i a_i -1$ corresponds to the charge (density) of the 
p-wave superconductor with the parameters $\mu=-2J_2$, $t=J_1$ and $\Delta=-J_1$ such that its derivative with respect to $J_2$ is the capacitance. We also have $\tau_{i-2}^x \tau_i^x =\tilde{\sigma}_{i-2}^y \tilde{\sigma}_i^y=-(a^{\dagger}_{i-2} a_i +h.c.) -(a^{\dagger}_{i-2} a^{\dagger}_i +h.c.)$. The roles of $J_1$ and $J_2$ are now inverted such that for the quantum Ising model, the topological phase
occurs for $J_2<J_1$ and the trivial phase for $J_2>J_1$. When $J_2<J_1$, the exponent $\frac{1}{4}$ \cite{Sachdev,Pfeuty} is associated to the long-range correlation of $\tau_i^x$ in the dual lattice \cite{Feng}. 

The operator $i c d$ in the $J_1-J_2$ model corresponds to a $\tau^y$ operator in the quantum Ising model on the dual lattice. 
This is another way to see why the short-range observable $\sigma_{2m}^y \sigma_{2m+1}^y$ on the $J_1-J_2$ lattice should be linked to the edge magnetic response associated to $\sigma_1^z=-i c_1 d_1$ when parameters are identical on the two lattices i.e. when $J_1=J_2$.
We show below how this logarithmic singularity at $J_1=J_2$ precisely encodes the edge response when adding the local magnetic field $h_1$. This gives rise to the
free energy correction $\Delta F = \int_0^{\beta} \langle \delta H(\tau)\delta H(0) \rangle d\tau$, with $\tau$ the imaginary time, 
where $\delta H=i h_1 c_1 d_1$. Then, this results in the edge spin susceptibility response
\begin{equation}
\label{chi1}
\chi_1 = \int_0^{\beta} \langle d_1(\tau)d_1(0)\rangle \times \langle c_1(\tau)c_1(0)\rangle d\tau.
\end{equation}
Since the $d_1$ fermion remains at zero energy $\langle d_1(\tau) d_1(0)\rangle = sgn(\tau)$. Within the correspondence onto the BCS theory, we can then evaluate
the Green's function in imaginary time for the Majorana fermion $c_1$.  At the quantum phase transition, when $J_1=J_2=J$ we find $\langle c_1(\tau) c_1(0)\rangle \approx \frac{\hbar}{\pi J\tau}$, such that \cite{SM}
\begin{equation}
\label{chi1}
\chi_1 \sim -\frac{1}{\pi}\frac{\hbar}{J}\ln |J_1-J_2|.
\end{equation}
The logarithmic singularity is reproduced with DMRG in Fig. \ref{susceptibility}. The linear behavior of $\langle \sigma_1^z\rangle$ shows a singular behavior when $J_1=J_2$ (red and blue
curves for $J_1<J_2$ and $J_2>J_1$). This form of $\chi_1$ draws an analogy with the 2CKM \cite{NozieresBlandin,EmeryKivelson,SenguptaGeorges,GiamarchiShraimanClarke} where the impurity spin-1/2 also decomposes itself naturally into two Majorana fermions at the Emery-Kivelson point, e.g. $c_1$ and $d_1$.
Note that if $d_1$ and $c_1$ would be bound as a spin-1/2 interacting with the continuum of states then $\chi_1$ goes to a fixed value at zero temperature \cite{SM}. In this way, the logarithmic singularity reveals the zero-energy Majorana fermion $d_1$. A relation with the 2CKM was reported numerically in the 2D Kitaev honeycomb spin model \cite{DasHoneycomb}. The edge spin susceptibility also takes a different form in the 2CKM with a 1D Luther-Emery superconducting wire showing Majorana edge fermions \cite{KarynEPL}. In the Supplementary Material, we also discuss the bulk responses \cite{SM}.

Above, we have presented short-range spin correlation functions and the local spin response at the edge. Here, we quantify the long-range resonance of the valence bonds through a quantum information probe, the bipartite fluctuations associated to the resonating valence bond observable $Q_{RVB}=\sum_{m\in A}\sigma_{2m-1}^x \sigma_{2m}^x$. The word bipartite means that we measure the fluctuations or variance (noise) associated to an observable on a region $A$ of the sample e.g. of size $l_A$ \cite{FluctuationsReview}. From the correspondence with the $\psi$-fermions, this is then similar to measure the bipartite charge fluctuations of the p-wave superconductor \cite{LoicKarynChristophe}. From Eq. (\ref{charge}), we identify the precise correspondence between 
the charge density in the superconductor $Q=\frac{1}{2}\sum_{m\in A}\psi^{\dagger}_m \sigma^z \psi_m = -\frac{1}{2}Q_{RVB}$ and the resonating valence bond observable $Q_{RVB}=\sum_{m\in A}\sigma_{2m-1}^x \sigma_{2m}^x$.
For the specific limit $|t|=|\Delta|$ of the p-wave superconducting wire (i.e. that is the situation for the $J_1-J_2$ model) then the bipartite charge fluctuations for the p-wave superconductor takes a simple {\it geometrical} form \cite{LoicKarynChristophe}
\begin{equation}
F_Q(A) = i_Q l_A + b \log l_A+{\cal O}(1)
\end{equation}
with $i_Q$ the density of charge fluctuations
\begin{equation}
\label{iQ}
i_Q = \hbox{lim}_{L\rightarrow +\infty} \frac{1}{L}(\langle Q^2\rangle - \langle Q\rangle^2) = \int_{-\pi}^{+\pi} \frac{dk}{4\pi}\sin^2 \theta_k.
\end{equation}
The charge fluctuations are introduced on the whole wire e.g. of length $L$ i.e. can also be interpreted as the quantum Fisher information which may be measured through dynamic susceptibilities \cite{Zoller}. 
Related to the quantum phase transition, the 2CKM also reveals an interesting charge dynamics \cite{ChristopheKaryn}.
The $\sin^2\theta_k$ function has a `local' geometrical meaning in $\theta_k$ space as $4A'_{\varphi}(\theta<\theta_c)A'_{\varphi}(\theta>\theta_c)$ \cite{KarynReview}.
The sub-leading negative $b$ coefficient generally traduces the presence of Majorana fermions \cite{LoicKarynChristophe}. Within the topological phase, $i_Q$ is globally related to the winding number
$m=\frac{1}{2\pi}\oint d\theta_k$ such that $i_Q= \frac{1}{4}m$. Bipartite fluctuations associated to $Q_{RVB}$ then measure $m$ in prefactor of the linear scaling function $l_A$. 
At the quantum phase transition, introducing $x=\frac{J_1}{J_2}=1\pm \epsilon$ (with $\epsilon\rightarrow 0$), then this results in a cusp \cite{LoicKarynChristophe} $\frac{\partial i_Q}{\partial x}\sim -i_Q(x=1^-)=-\frac{1}{4}$. 

To summarize, we have introduced short-range spin correlations and the local magnetization at the edge to reveal the topological invariant and the topological phase transition in the $J_1-J_2$ quantum spin model, through their derivatives. 
The edge magnetization is a physical marker of the topological invariant and of the Majorana zero mode(s). 
We have presented a correspondence between the resonance of valence bonds and bipartite charge fluctuations in the p-wave superconductor. 
This work illustrates the usefulness of measuring the capacitance of a p-wave superconducting wire \cite{Aghaee,FrederickKaryn}. The generalization of the Hamiltonian with additional Ising interactions
allows us to propose the engineering of this model and of zero-energy Majorana fermions \cite{SM} through realistic quantum circuits \cite{HH,KarynReview}; we also provide a physical insight \cite{SM} onto realizations of AKLT models \cite{AKLT}.
We hope that this platform may acquire similar attention as the quantum Ising model \cite{DeGennes,Pfeuty,XMi} and the Kitaev p-wave superconducting wire \cite{Kitaev}. For the 2D Kitaev honeycomb spin model \cite{Kitaevhoneycomb}, important efforts also exist to relate local and dynamical observables with the quantum phase transitions and entanglement entropy \cite{Wang,KangWang,Trivedi}. Analytical relations between short-range spin observables and the topological invariant 
remain to be addressed in 2D.

Magali Korolev acknowledges Ecole Polytechnique for the financial support of her PhD thesis. Fan Yang was supported by the Knut and Alice Wallenberg Foundation (KAW) via the project Dynamic Quantum Matter (2019.0068).

\title{Supplementary Material}

\section{Generalization of the Hamiltonian through the AKLT model}

{\it Trivial Phase:} Suppose we are within the {\it trivial phase} where $J_1\gg J_2$. Physical results will remain very similar for observables if we include an additional Ising term along $z$ direction $J_{1}^z$ for the strong bond $\{j;j+1\}$ such that 
$(J_1,J_{1}^z)\gg J_2$. Including e.g. $J_{1}^z\sigma_j^z \sigma_{j+1}^z$ this results in the Hamiltonian
\begin{equation}
\label{Isinginteractions}
{\cal H}=\sum_{j=2m-1} (-i)c_j c_{j+1}(J_1-i J_{1}^z d_j d_{j+1}) +iJ_2 c_{j+1} c_{j+2}.
\end{equation}
As long as $(J_1,J_{1}^z)\gg J_2$, the ground state $|\psi\rangle=\prod_j |\psi_{j;j+1}\rangle$ is unique. We have justified in the Letter that for $J_{1}^z=0$ the presence of free Majorana fermions $d_j$, $d_{j+1}$ corresponds to a $\ln 2$ entropy 
associated to the classical degeneracy of equiprobable ground states of the Ising model $\uparrow_x\downarrow_x$ and $\downarrow_x\uparrow_x$ on a link $\{j;j+1\}$. When $J_{1}^z=0=J_2$, the model is classical and similarly for the equiprobable
eigenstates. If we re-write the information along $z$ direction, $\uparrow_x \downarrow_x = \frac{1}{2}(\uparrow_z + \downarrow_z)\otimes (-\downarrow_z +\uparrow_z)$ and 
$\downarrow_x \uparrow_x = \frac{1}{2}(-\downarrow_z + \uparrow_z)\otimes (\uparrow_z+\downarrow_z)$. For $J_{1}^z\neq 0$, the superposition of these two eigenstates then leads to
\begin{equation}
|\psi\rangle_{j;j+1} = \frac{1}{\sqrt{2}}(-\uparrow_z \downarrow_z + \downarrow_z \uparrow_z)=\frac{1}{\sqrt{2}}(-\uparrow_x \downarrow_x + \downarrow_x \uparrow_x)
\end{equation}
such that on a link $\{j;j+1\}$, the ground state energy is $-J_1-J_{1}^z$. For the second equality, we have multiplied the eigenstate by a global phase of $\pi$ that will not modify the result on correlation functions
\begin{equation}
\sigma_j^z \sigma_{j+1}^z |\psi_{j;j+1}\rangle = \sigma_j^x \sigma_{j+1}^x |\psi_{j;j+1}\rangle = - |\psi_{j;j+1}\rangle.
\label{correlation}
\end{equation}
When $(J_2,J_{2}^z)\rightarrow 0$, the ground state is a collection of singlet valence bonds i.e. a specific example of gapped spin liquid. In terms of Majorana fermions, Eqs. (\ref{correlation}) lead to
\begin{equation}
i c_j c_{j+1}|\psi\rangle = i d_{j+1} d_j |\psi\rangle = +|\psi\rangle.
\end{equation}
This is then equivalent to modify $J_1\rightarrow J_1+J_{1}^z$ in Eq. (\ref{Isinginteractions}) maintaining the same form of Hamiltonian for the $c$-Majorana fermions through $E(k)$; e.g. there is a corresponding term in $2J_1 J_2 \cos(kl)$ with a doubled lattice spacing $l=2a$. As we show below for the topological phase, a link $J_2$ equally admits a singlet valence bond e.g. along $z$ axis. Therefore, singlet
states are allowed to resonate between different adjacent bonds. The phase is a gapped quantum spin liquid i.e. $\langle \sigma_j^{\mu}\rangle=0$ with $\mu=x,y,z$. As long as $J_{2}^z=0$, since $i d_j d_{j+1}$ commutes with the Hamiltonian this implies that 
even if the ground state wavefunction becomes modified with $J_2$ then the parity operator $i d_j d_{j+1}$ will preserve the identity $i d_{j+1} d_j |\psi\rangle = +|\psi\rangle$. This shows that it is yet possible to identify $\langle i d_{j} d_{j+1}\rangle =-1$
in the Hamiltonian when approaching the transition very closely. If we add a term $J_{2}^z\ll J_{1}^z$ since $i d_{j+1} d_j |\psi\rangle_{j;j+1} = -1|\psi\rangle_{j;j+1}$ this will maintain $\langle i d_{j+1} d_{j+2}\rangle=0$ as if $J_{2}^z=0$.  

\begin{figure}[t]
\includegraphics[width=8.5cm]{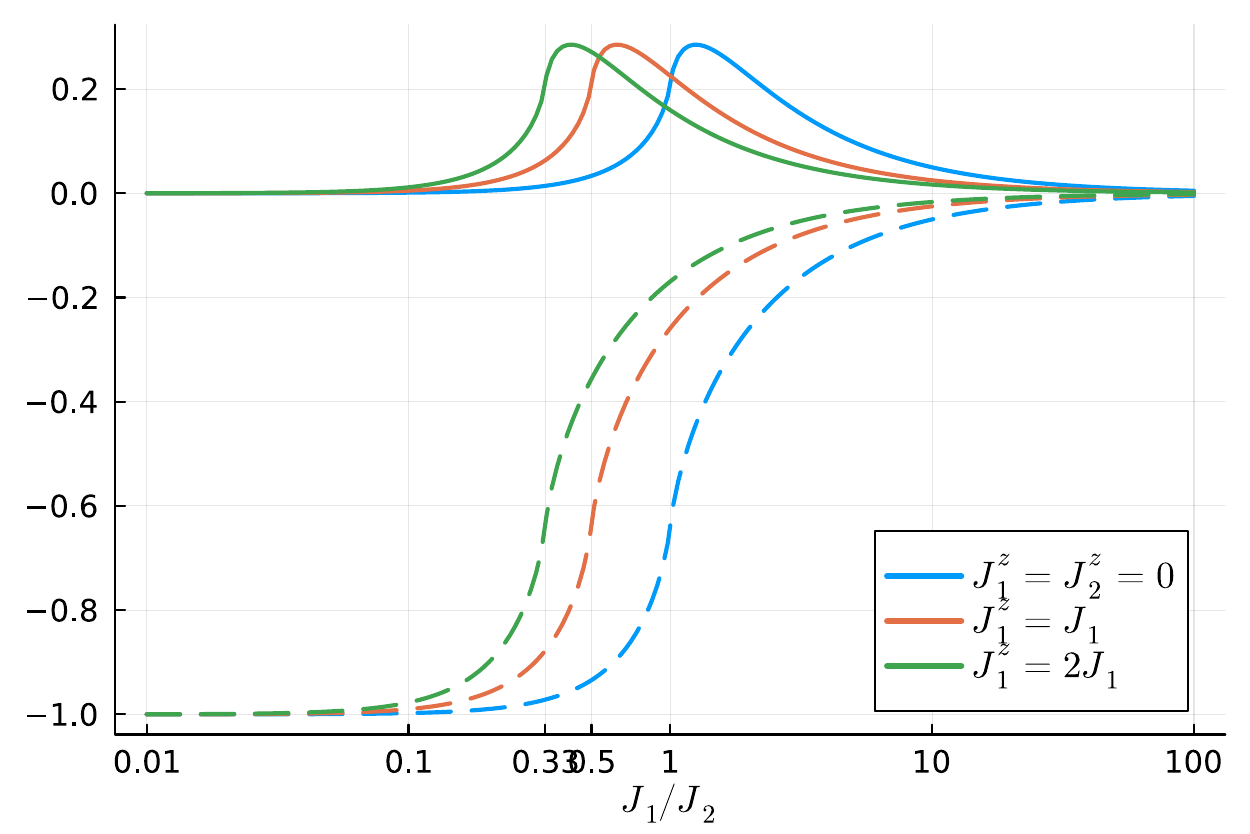}
\includegraphics[width=8.5cm]{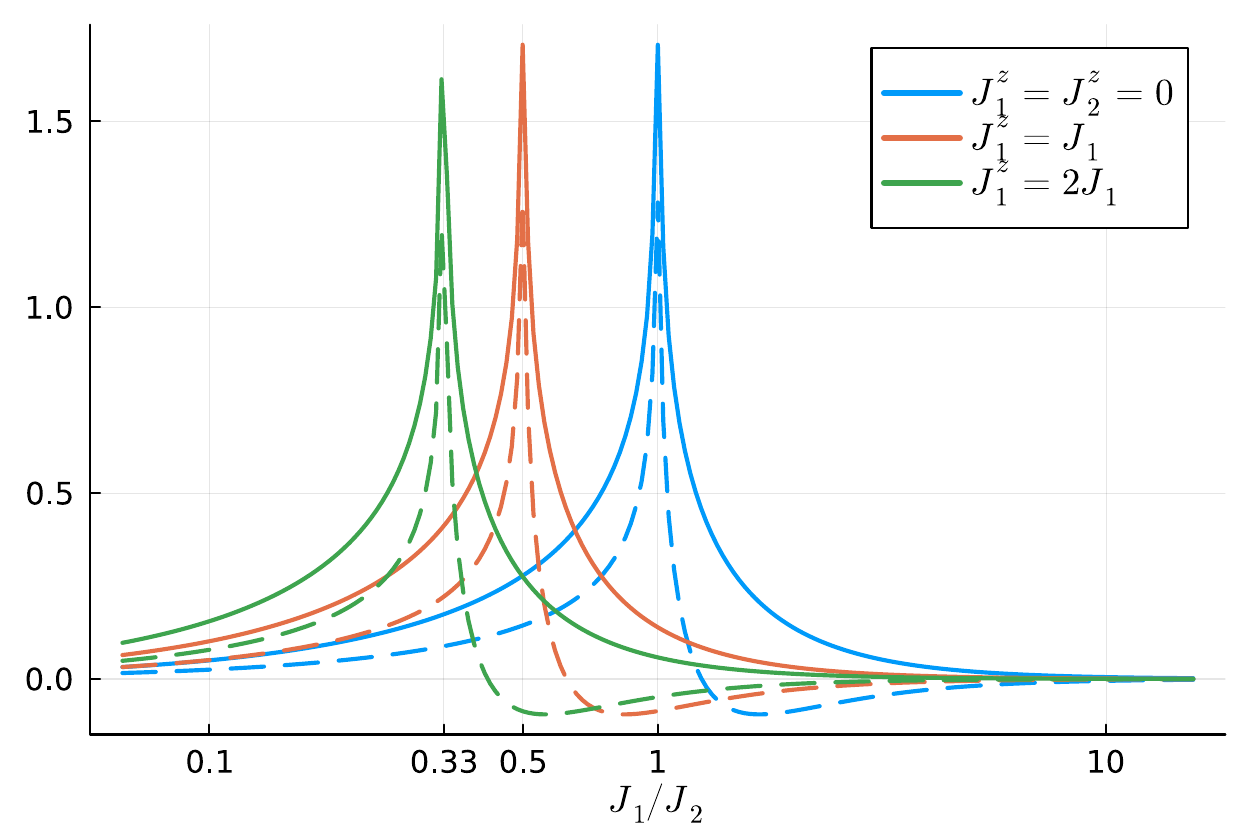}
\caption{(Top) $\langle \sigma_{2m-1}^x\sigma_{2m}^z\sigma_{2m+1}^z\sigma_{2m+2}^x\rangle$ (solid lines) and $\langle \sigma_{2m}^y\sigma_{2m+1}^y\rangle$ (dashed lines)
from DMRG when including an additional term $J_{1}^z$. A link to the DMRG data can be found in Ref. [33]
of the Letter. (Bottom) Logarithmic divergences of the derivatives associated to short-range spin correlation functions.}
\label{J1zfigure}
\end{figure}

When including $J_{1}^z$ in the analysis, through the DMRG approach we find e.g. that the quantum phase transition occurs precisely when  $J_1+J_{1}^z=J_2$ even for $J_1^z=2J_1$; see Figures \ref{J1zfigure}. 
The form of the logarithmic divergences in the derivatives of observables agree with Eq. (20) in the Letter. 

Interestingly, we may also include a term $J_{1}^y\sigma_j^y \sigma_{j+1}^y$ that turns into $+J_{1}^y i d_j d_{j+1}$; the eigenvalue $i d_j d_{j+1}=-1$ leads to the lowering of energy by an amount $-J_{1}^y$.
This will be relevant for practical applications related to the platform introduced in Ref. [1] of the Supplementary Material references list. See Section hereafter. We verify with DMRG that the transition remains localized at $J_1=J_2$ when including
the additional $J_1^y$ term (only) i.e. fixing $J_1^z=0$. When transforming the interaction $J_1^x$ as an $xy$ interaction $r_{xy}\sigma_j^+ \sigma_{j+1}^- +h.c.$, corresponding to $J_1^x=J_1^y$, then we
verify through DMRG that results remain identical and that the entanglement entropy remains equal to $c=\frac{1}{2}$ at the transition. Modifying the interactions $J_1=J_1^x$ and $J_2=J_2^y$ as $xy$ interactions will be
interesting for applications, see next Section.
 
\begin{figure}[t]
\includegraphics[width=8.5cm]{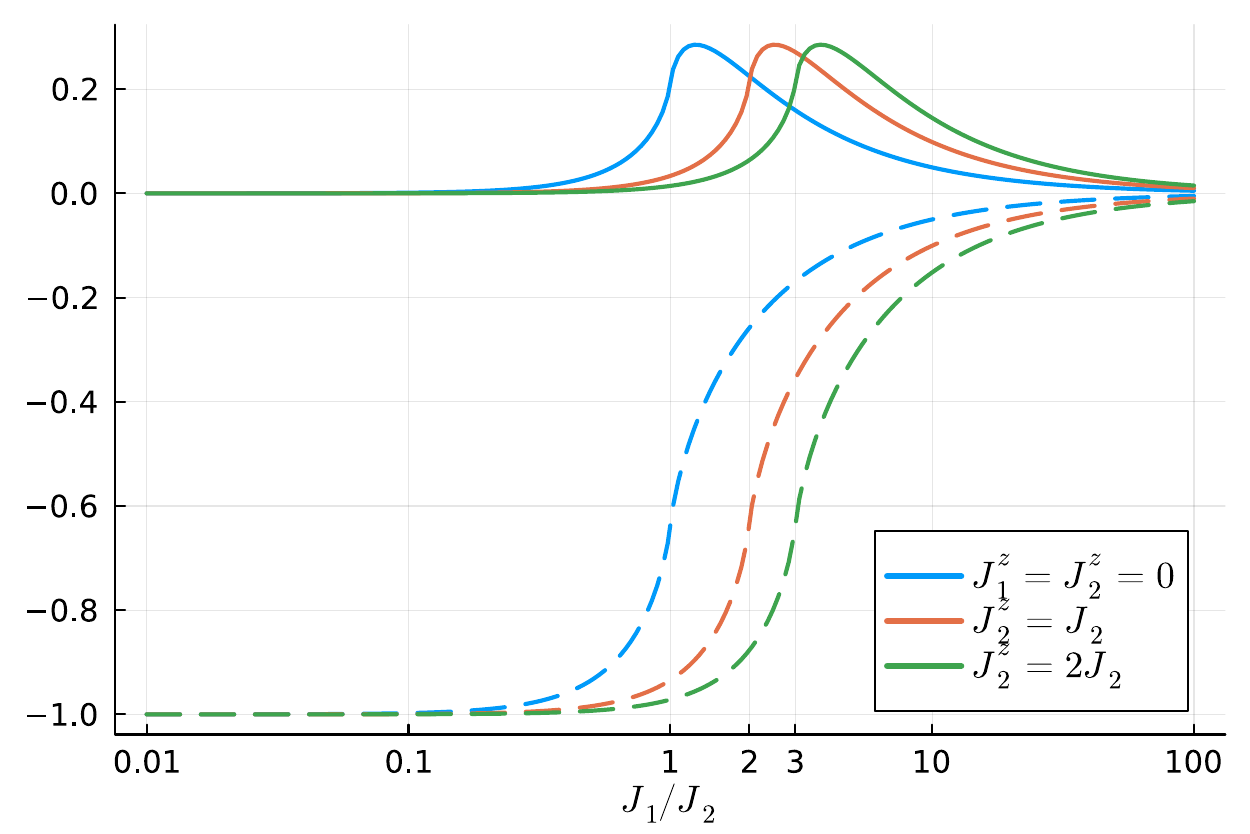}
\includegraphics[width=8.5cm]{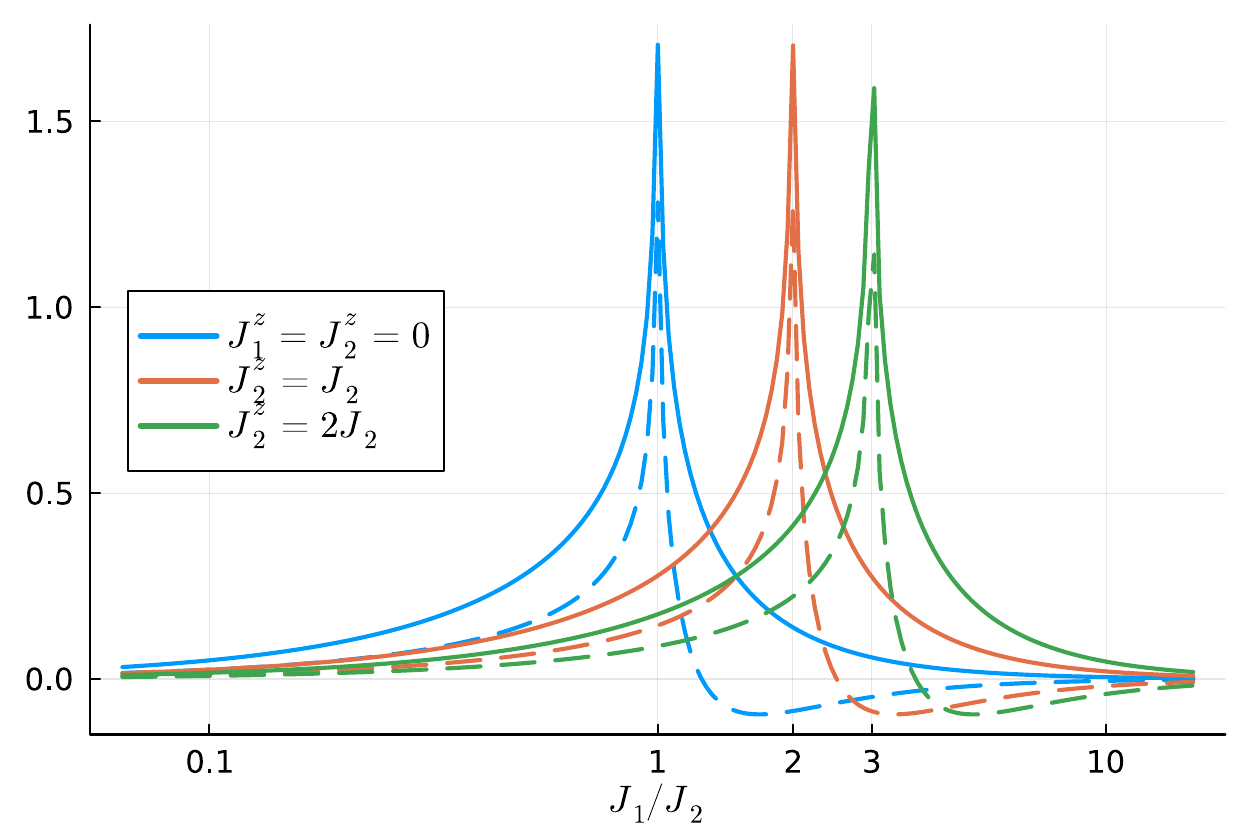}
\caption{(Top) $\langle \sigma_{2m-1}^x\sigma_{2m}^z\sigma_{2m+1}^z\sigma_{2m+2}^x\rangle$ (solid lines) and $\langle \sigma_{2m}^y\sigma_{2m+1}^y\rangle$ (dashed lines) from DMRG when including an additional term $J_{2}^z$. 
A link to the DMRG data can be found in Ref. [33] of the Letter. (Bottom) Logarithmic divergences of the derivatives associated to short-range spin correlation functions.}
\label{J2z}
\end{figure}

{\it Topological Phase:} If we are within the {\it topological phase} i.e. $J_2\gg J_1$ then the results stay similar as long as $(J_2,J_{2}^z)\gg J_1$. Indeed, in that case,
\begin{equation}
J_{2}^z\sigma_{j+1}^z \sigma_{j+2}^z = J_{2}^z (i c_{j+1} d_{j+1})(i d_{j+2} c_{j+2}).
\label{J2zcoupling}
\end{equation}
In the presence of $J_2$ and $J_{2}^z$ the ground state is also unique (non-degenerate) $|\psi\rangle' = \prod_{j} |\psi\rangle_{j+1;j+2}$ such that
\begin{equation}
\label{identities}
i c_{j+1} c_{j+2} |\psi'\rangle = i d_{j+2} d_{j+1} |\psi'\rangle = - |\psi'\rangle.
\end{equation}
We can precisely re-write the two classical ground states found for $J_{2}^z=0$ as $\uparrow_y \downarrow_y = \frac{1}{2}(\uparrow_z + i \downarrow_z)\otimes (-\uparrow_z +i\downarrow_z)$
and $\downarrow_y \uparrow_y = \frac{1}{2}(-\uparrow_z + i \downarrow_z)\otimes (\uparrow_z +i\downarrow_z)$. For any $J_{2}^z\neq 0$, the unique ground state corresponds to the specific superposition
\begin{equation}
|\psi\rangle_{j+1;j+2} = \frac{+i}{\sqrt{2}}(\uparrow_z \downarrow_z - \downarrow_z \uparrow_z) = \frac{1}{\sqrt{2}}(\uparrow_y \downarrow_y - \downarrow_y \uparrow_y).
\end{equation}
In this way, 
\begin{equation}
\sigma_{j+1}^z \sigma_{j+2}^z |\psi_{j+1;j+2}\rangle = \sigma_{j+1}^y \sigma_{j+2}^y |\psi_{j+1;j+2}\rangle = - |\psi_{j+1;j+2}\rangle.
\end{equation}
The form of the Hamiltonian in terms of the $c$-fermions remains also identical modulo the transformation $J_{2}\rightarrow J_{2}+J_{2}^z$. This fact should again remain true when approaching the transition point.
We remind that within the topological phase,  we have a nice analytical equation relating observable and topological invariant, $\langle\sigma_{j+1}^y \sigma_{j+2}^y\rangle=-{\cal C}$.
We can then add one sentence to the proof: If we add a term $J_{1}^z\ll (J_{2}^z,J_2)$ since $i d_{j+2} d_{j+1} |\psi'\rangle_{j+1;j+2} = -1|\psi\rangle_{j;j+1}$ this will maintain $\langle i d_{j} d_{j+1}\rangle=0$
as if $J_{1}^z=0$. This justifies why the edge magnetization and local magnetic susceptibility keep a similar forms as in Fig. 1 of the Letter. The local spin magnetization at an edge then represents a robust local order
parameter for the topological phase transition. We verify this important conclusion through DMRG. When including $J_{2}^z$ in the analysis, the DMRG approach also reproduces well that the quantum phase transition occurs when  $J_2+J_{2}^z=J_1$; 
see Figures \ref{J2z}. The form of the logarithmic divergences in the derivatives of observables also agree with Eq. (20) in the Letter.
 
{\it Quantum Phase Transition:} Here, we discuss properties of the quantum phase transition when including both $J_1^z$ on a bond $\{j;j+1\}$ and $J_2^z$ on a bond $\{j+1;j+2\}$ with $j=2m-1$.
The origin of quantum fluctuations at the quantum phase transition comes from the fact that the parity operators $i c_j c_{j+1}$ and $i c_{j+1} c_{j+2}$ do not commute. 
At the quantum phase transition $J_1=J_2$ and $J_1^z=J_{2}^z$, then the whole model takes the form
\begin{eqnarray}
{\cal H} = \sum_{j=2m-1} && (-i) c_j c_{j+1}(J_1 - i J_{1}^z d_j d_{j+1})  \nonumber \\
&+& i c_{j+1}c_{j+2}(J_1 + i J_{1}^z d_{j+1} d_{j+2}).
\end{eqnarray}
To find the universality class of the quantum phase transition we apply the Taylor formula $c_{j+2} = c_j +2l \partial_j c_j$ with $l$ the lattice spacing. We can then allow for a similar formula
in terms of the fermion $d_{j+2}$ such that
\begin{eqnarray}
\hskip -0.6cm {\cal H} = \sum_{j=2m-1} (-i) J_1 c_j c_{j+1} + J_{1}^z (-i) c_j c_{j+1}(-i) d_j d_{j+1}  \nonumber \\
+ i c_{j+1}(c_j+2l\partial_j c_j)(J_1 + i J_{1}^z d_{j+1}(d_j +2l \partial_j d_j)).
\end{eqnarray}
Suppose we minimize first the energy of the terms associated to the links $\{j;j+1\}$ with $i c_j c_{j+1}|\psi_{j;j+1}\rangle=+1=i d_{j+1} d_j|\psi_{j;j+1}\rangle$ then we obtain quantum fluctuations from the Taylor formula associated to the links $\{j+1;j+2\}$. The dominant
terms in those zero-energy fluctuations are
\begin{equation}
{\cal H}' = \sum_{j=2m-1} i(J_1+J_{1}^z) c_{j+1} 2l \partial_j c_j  +  i J_{2}^z d_{j+1} 2l \partial_j d_j.
\end{equation}
In the last formula, we have re-instored $J_{2}^z$ which is equal to $J_{1}^z$ at the transition. We can proceed as in the Letter and justify that when $J_{2}^z\neq 0$ each Majorana sector produces a quantum liquid associated to two 2D classical Ising models or two central charges $c=\frac{1}{2}$ (corresponding to a total central charge of one) in the entanglement entropy. As long as $J_{1}^z\neq 0$, we obtain a symmetric result if we first fix the energy of the terms associated to the links $\{j+1;j+2\}$. 

\section{Realization with a wire of resonating Bloch spheres}

To realize the model, we begin with {\it a pair of spins} described through the Hamiltonian [1]
\begin{equation}
\label{spinsmodel}
{\cal H}_{j;j+1} = - {\bf H}_1\cdot \mathbfit{\sigma}_j - {\bf H}_1\cdot \mathbfit{\sigma}_{j+1} +J_{1}^z \sigma_j^z \sigma_{j+1}^z.
\end{equation}
The two spins present inversion symmetry
\begin{equation}
{\bf H}_1 = (H_1\sin \theta \cos \varphi, H_1 \sin \theta \sin \varphi, H_1\cos\theta+M_1).
\end{equation}
It is worth mentioning that the two-spins model with radial magnetic fields is realized in Ref. [2] in quantum circuits related to the model of Ref. [1] in Eq. (\ref{spinsmodel}).

We can first fix parameters such that close to south pole a physical situation identical to the $(J_1,J_{1}^z)$ model occurs. This requires to adjust $H_1-M_1\ll J_{1}^z\ll H_1+M_1$.
Close to the south pole $\theta=\pi$, the effective Hamiltonian then takes the form [1]
\begin{equation}
{\cal H}_{eff,j;j+1} = J_{1}^z \sigma_j^z \sigma_{j+1}^z -\frac{H_1^2 \sin^2 \theta}{J_{1}^z} \sigma_j^x \sigma_{j+1}^x.
\label{interactionx}
\end{equation}
The term $J_1$ comes from second-order perturbation theory on a bond $\{j;j+1\}$ where the perturbative Hamiltonian is the transverse field e.g. $-H_1\sin \theta (\sigma_j^x + \sigma_{j+1}^x)$.
Higher-order terms do not modify the structure of the effective Hamiltonian for a pair of adjacent sites.
The polar angle can be adjusted when navigating in time $\theta=vt$, as realized [2], to measure geometrical and topological properties, and we can also navigate on a line of fixed azimuthal angle $\varphi$ such that e.g. $\varphi=0$ or $\varphi=\frac{\pi}{2}$. If $\varphi=0$ this will favor an Ising interaction along $x$ as in the Equation above. If $\varphi=\frac{\pi}{2}$ then the induced Ising interaction will be along $y$ direction, as discussed below for the pair $\{j+1;j+2\}$. 
We propose then to adjust the angle $\theta=\theta^*$ close to $\theta=\pi$, producing an additional ferromagnetic coupling $J_1=-\frac{H_1^2 \sin^2 \theta}{J_{1}^z}$. This is also judicious to discuss these perturbations because in an experiment, it is appropriate
to understand the role of transverse fields in the Hamiltonian.  After reaching this polar angle value, we maintain this angle fixed i.e. $vt\rightarrow \theta^*$. 

On a link $\{j;j+1\}$ with $j=2m-1$, the ground state takes the form $|\psi\rangle_{j;j+1}=\frac{1}{\sqrt{2}}(\uparrow_z \downarrow_z + \downarrow_z \uparrow_z)$. Along $x$ axis, it takes the form $|\psi\rangle_{j;j+1}=\frac{1}{\sqrt{2}}(\uparrow_x \uparrow_x - \downarrow_x \downarrow_x)$. 
The ground state energy is $J_1-J_1^z$. The formation of the entangled state around south pole can be observed through a pair of fractional
$\frac{1}{2}$ topological numbers [1,3,4]. Measuring the fractional topological number for one spin is a way to show the existence or formation of
the entangled wave-function around south pole.  In terms of Majorana fermions $c$ and $d$, if we flip the sign of $J_1$ to have a ferromagnetic interaction $J_1<0$, this implies a modification of Eq. (\ref{identities}) such that
\begin{equation}
i c_j c_{j+1} |\psi\rangle = - |\psi\rangle = - id_j d_{j+1} |\psi\rangle.
\end{equation}
The Hamiltonian then is equivalent to
\begin{equation}
{\cal H}_{j;j+1} = (-i) c_j c_{j+1}(J_1-i J_1^z d_j d_{j+1}).
\end{equation}
We emphasize here that a direct ferromagnetic Ising term along $y$ direction coming from $r_{xy} \sigma_j^+ \sigma_{j+1}^-+h.c.$ with $r_{xy}<0$ would result in an additional
term of the form $r_{xy} (i d_j d_{j+1})$ which can then lower the energy since ${\cal H}_{eff,j;j+1}$ favors  $id_j d_{j+1} |\psi\rangle=+|\psi\rangle$.
For any pair $\{j;j+1\}$ with $j=2m-1$, we can engineer in principle a similar Hamiltonian. The direct interaction $r_{xy}$ would help us maintaining e.g. $J_1\gg J_1^z$ and at the same time
this would not induce additional interaction terms from perturbation theory between pairs $\{j;j+1\}$ and $\{j+2;j+3\}$, i.e. we can maintain $\theta=\theta^*$ closer to $\theta=\pi$.

For the pair $\{j+1;j+2\}$ we can proceed in a similar way and prepare the system with $\varphi=\frac{\pi}{2}$. In this way,
around south pole, we obtain the effective Hamiltonian
\begin{equation}
{\cal H}_{eff,j+1;j+2} = J_{2}^z \sigma_j^z \sigma_{j+1}^z -\frac{H_2^2 \sin^2 \theta}{J_{2}^z} \sigma_j^y \sigma_{j+1}^y.
\end{equation}
The interaction $J_{2}=-\frac{H_2^2 \sin^2 \theta}{J_{2}^z}$ is also ferromagnetic. The corresponding ground state then is $|\psi'\rangle_{j+1;j+2} = \frac{1}{\sqrt{2}}(\uparrow_y \uparrow_y - \downarrow_y \downarrow_y)$ which can be re-written as
$|\psi'\rangle = \frac{i}{\sqrt{2}}(\uparrow_z \downarrow_z + \downarrow_z \uparrow_z)$. The ground state energy is $J_2-J_{2}^z$. It is then practical to also allow for a direct ferromagnetic $xy$ interaction [1-3] that will reinforce the role of the ferromagnetic
$J_2=J_2^y$ spin coupling (here, along $y$ direction), reinforcing the ferromagnetic parameter $J_2$ i.e. allowing us to reach safely e.g. the limit $|J_2|\gg J_{2}^z$. The coupling $J_2^x \sigma_{j+1}^x \sigma_{j+2}^x$ would produce
a term coupling the fermions $d$ as $-i J_2^x d_{j+1} d_{j+2}$, which indeed reinforces the role of the coupling $J_2^z$ in Eq. (\ref{J2zcoupling}) through the energetic minimization $\langle -i d_{j+1} d_{j+2} \rangle =+1$.

It should be emphasized that for the specific situation, $J_{1}^z=J_{2}^z$ and $J_1=J_2$ then the model shows a fractional topological number $\frac{1}{2}$ for each sphere within the thermodynamical limit [1,4]. In the Letter, we have justified that the quantum phase transition in the $J_1-J_2$ model e.g. can be thought of as a metal of Majorana fermions with central charge $c=\frac{1}{2}$. The geometrical description relates then different important characterizations, the central charge from the quantum field theory and a fractional topological number which e.g. measures the formation of entangled wave-functions at one pole. The occurrence of such fractional numbers requires an inversion symmetry between all spins therefore when we slightly modify e.g. $(J_2, J_{2}^z)$ this justifies
the occurrence of integer topological numbers, one or zero, in the model i.e. when $(J_2,J_2^z)\gg (J_1,J_1^z)$ or when $(J_2,J_2^z)\ll (J_1,J_1^z)$. 

Alternatively, we may realize the Hamiltonian with charge qubits where charge $0$ and $1$ can describe the pseudo spin-1/2 on each site. A capacitance between a pair of quantum dots then describes an Ising coupling along $z$ direction and
an $r_{xy}$ coupling may then correspond to a hopping term between a pair of quantum dots. Adjusting the distance between two charge islands may allow us to engineer different capacitance terms and hopping terms.

\section{Diagonalization of Hamiltonian and BCS ground state}

To describe the topological properties of this system and the ground-state wavefunction it is then useful to introduce the Nambu or Nambu-Jona-Lasinio representation in momentum space $\Psi_k^{\dagger}=(\psi^{\dagger}_k,\psi_{-k})$ and to write the 
Bogoliubov-de Gennes Hamiltonian as ${\cal H}=\sum_k \Psi^{\dagger}_k {\cal H}(k) \Psi_k$ with ${\cal H}(k)$ a $2\times 2$ matrix associated to a spin-$\frac{1}{2}$ particle on the Bloch sphere
\begin{equation}
{\cal H}(k) = \begin{pmatrix}
\xi_k & \Delta_k \\
\Delta_k^* & -\xi_k
\end{pmatrix}
\end{equation}
where
\begin{eqnarray}
\label{parameters}
\xi_k &=& -J_1-J_2 \cos (kl) \\ \nonumber
\Delta_k &=& -iJ_2\sin(k l).
\end{eqnarray}
The eigen-energies are $E_{\pm}(k)=\pm E(k)$ with
\begin{equation}
E(k)=\sqrt{J_1^2+J_2^2+2J_1 J_2\cos(kl)}.
\end{equation}

Writting the $2\times 2$ matrix in this way, the associated Brillouin zone for the fermions $\psi$ takes the usual form $k\in [-\frac{\pi}{l};\frac{\pi}{l}]$ even though the spin model has a periodicity of $2a$. 
The ground-state wave function takes the BCS form
\begin{equation}
\label{BCS}
|BCS\rangle = {\cal R}_0 \prod_{k>0}^{\pi/l} \left(\sin \frac{\theta_k}{2} + i \cos\frac{\theta_k}{2} \psi^{\dagger}_k \psi^{\dagger}_{-k}\right) {\cal R}_{\pi}|0\rangle
\end{equation}
with ${\cal R}_0$ and ${\cal R}_{\pi}$ taking into account that the gap is closing at $k=0$ and $kl=\pi$, i.e. ${\cal R}_0|0\rangle=\psi^{\dagger}_0|0\rangle$ if $J_1>-J_2$ and to $|0\rangle$ otherwise, and
${\cal R}_{\pi}|0\rangle=\psi^{\dagger}_{\pi}|0\rangle$ if $J_1>J_2$ and to $|0\rangle$ otherwise. Since $(J_1,J_2)>0$ the gap closing cannot happen at $kl=0$.
Topological properties of the superconducting wire can be elegantly described on the (surface of the) Bloch sphere of the spin-$\frac{1}{2}$
particle \cite{KarynReview} through the correspondences 
\begin{eqnarray}
\cos\theta_k &=& \frac{J_1+J_2\cos kl}{E(k)} \\ \nonumber
\sin\theta_k e^{-i\varphi} &=& -ie^{i\phi}\frac{J_2\sin(k l)}{E(k)}.
\end{eqnarray}
Here, $\theta_k$ and $\varphi_k$ are the polar and azimuthal angles on the sphere.
We underline that $\theta$ can be mapped onto a function of the wave-vector $k$ whereas the azimuthal angle $\varphi$ on the sphere corresponds in fact to the superfluid phase associated to the order parameter $\Delta$. 
If we introduce precisely a phase $\phi+\pi$ to $\Delta$ this is equivalent to modify $\psi^{\dagger}_k \psi^{\dagger}_{-k}\rightarrow -e^{i \phi}\psi^{\dagger}_k \psi^{\dagger}_{-k}$. Then, we fix the azimuthal angle on the sphere such that $e^{-i\varphi}=-i e^{ i\phi}$.
Compared to usual definitions, since we are in a situation with $t=-\Delta$ we can allow for a gauge choice for the phase as $\phi+\pi$ to absorb the relative sign between $t$ and $\Delta$. The quasiparticle operators take the form 
\begin{equation}
\eta_k = \cos\frac{\theta_k}{2}\psi_k + i\sin\frac{\theta_k}{2}\psi^{\dagger}_{-k}.
\end{equation}
The Hamiltonian takes the precise form ${\cal H}(k) = \sum_{k=-\frac{\pi}{l}}^{k=\frac{\pi}{l}} -E(k)(\eta^{\dagger}_{k} \eta_k - \eta_{-k}\eta_{-k}^{\dagger})$ such that the ground satisfies $\langle BCS|\eta^{\dagger}_k \eta_k|BCS\rangle=1$. 
Introducing the phase $\phi+\pi$ to the fermions then modifies the $S^y$ operator as $S^y = i (e^{i\phi}\psi^{\dagger}_k \psi^{\dagger}_{-k} - e^{-i\phi}\psi_{-k}\psi_k)$ such that
\begin{equation}
\langle BCS| S^y| BCS\rangle = - \sin \theta_k.
\end{equation}
Eq. (\ref{BCS}) is modified as
\begin{equation}
|BCS\rangle = {\cal R}_0 \prod_{k>0}^{\pi/l} \left(\sin \frac{\theta_k}{2} - i e^{i\phi}\cos\frac{\theta_k}{2} \psi^{\dagger}_k \psi^{\dagger}_{-k}\right) {\cal R}_{\pi}|0\rangle
\end{equation}

\section{Evaluation of the Local Spin Susceptibility}

Here, we aim to evaluate the edge magnetic susceptibility
\begin{equation}
\frac{\chi+\chi^*}{2} = Re(\chi)=\int_0^{\beta} d\tau \langle c_1(\tau)c_1(0)\rangle \times \langle d_1(\tau) d_1(0)\rangle.
\end{equation}
The fermion $d_1$ is free at zero-energy such that $d_1(\tau)=d_1(0)$ and
\begin{equation}
\langle d_1(\tau)d_1(0)\rangle = sgn(\tau).
\end{equation}
The local edge susceptibility is then evaluated from the Green's function of the fermion $c_1$: $\langle c_1(\tau) c_1(0)\rangle$. From the mapping onto the bond fermions, this can be evaluated through the identification
\begin{eqnarray}
\label{formulac1}
\langle c_1(\tau) c_1(0)\rangle &=& \langle \psi_1(\tau)\psi_1(0)\rangle + \langle \psi_1^{\dagger}(\tau) \psi_1^{\dagger}(0)\rangle \\ \nonumber
&+& \langle \psi_1^{\dagger}(\tau)\psi_1(0)\rangle + \langle \psi_1(\tau)\psi_1^{\dagger}(0)\rangle,
\end{eqnarray}
close to the quantum phase transition, in imaginary time. To account for the time-dependence of the electron operators, we can re-write the fermion operators in terms of the quasiparticles operators
\begin{eqnarray}
\psi_k = \cos\frac{\theta_k}{2}\eta_k - i \sin\frac{\theta_k}{2}\eta_{-k}^{\dagger}.
\end{eqnarray}
We also have 
\begin{equation}
\psi_1 = \frac{1}{\sqrt{M}}\sum_k e^{i k}\psi_k.
\end{equation}
The lattice spacing $l=1$ is set to unity in this Section. 
The Hamiltonian for the quasiparticles read
\begin{equation}
{\cal H} = \sum_{k=-\pi}^{k=+\pi} (-E(k))(\eta_k^{\dagger} \eta_k - \eta_{-k}\eta_{-k}^{\dagger})
\end{equation}
The BCS ground state is then defined such that $\langle \eta_k^{\dagger}\eta_k\rangle = 1$ and $\langle \eta_{-k}^{\dagger} \eta_{-k}\rangle = 1$ at zero temperature. The time-dependence of $\eta_k$ quasiparticles can be clarified from Heisenberg equations of motion
\begin{equation}
\frac{d\eta_k}{dt} = \frac{i}{\hbar}[{\cal H},\eta_k] = \frac{i}{\hbar} E(k) \eta_k.
\end{equation}
This results in
\begin{eqnarray}
 \eta_{\pm k}(t) =  \eta_{\pm k}(0) e^{\frac{i E(k) t }{\hbar}} \\ \nonumber
 \eta^{\dagger}_{\pm}(t) = \eta^{\dagger}_{\pm k}(0) e^{-\frac{i E(k) t }{\hbar}}.
\end{eqnarray}
In imaginary time, we apply the transform $\tau=it$ or $t=-i\tau$.
We can verify that the two pairing terms in Eq. (\ref{formulac1}) gives an {\it imaginary} response
\begin{equation}
\langle \psi_1(\tau) \psi_1(0)\rangle = -\frac{i}{M} \sum_k \sin\frac{\theta_k}{2} \cos\frac{\theta_k}{2} e^{-\frac{E(k)\tau}{\hbar}}.
\end{equation}
We can also verify that
\begin{equation}
\langle \psi_1(\tau)\psi_1(0)\rangle^* = - \langle \psi_1^{\dagger}(0) \psi_1^{\dagger}(\tau)\rangle.
\end{equation}

Then, we find
\begin{equation}
\langle \psi_1^{\dagger}(\tau)\psi_1(0)\rangle = \frac{1}{M}\sum_{k=-\pi}^{+\pi} \cos^2\frac{\theta_k}{2} e^{-\frac{E(k)\tau}{\hbar}} \langle \eta^{\dagger}_k(0)\eta_k(0)\rangle.
\end{equation}
This simplifies into
\begin{equation}
\langle \psi_1^{\dagger}(\tau)\psi_1(0)\rangle = \frac{1}{M}\sum_{k=-\pi}^{+\pi} \cos^2\frac{\theta_k}{2} e^{-\frac{E(k)\tau}{\hbar}}.
\end{equation}
In a similar way, we identify
\begin{equation}
\langle \psi_1(\tau)\psi_1^{\dagger}(0)\rangle = \frac{1}{M}\sum_{k=-\pi}^{+\pi} \sin^2\frac{\theta_k}{2} e^{-\frac{E(k)\tau}{\hbar}} \langle \eta^{\dagger}_{-k}(0)\eta_{-k}(0)\rangle,
\end{equation}
such that
\begin{equation}
\langle \psi_1(\tau)\psi_1^{\dagger}(0)\rangle = \frac{1}{M}\sum_{k=-\pi}^{+\pi} \sin^2\frac{\theta_k}{2} e^{-\frac{E(k)\tau}{\hbar}}.
\end{equation}
Therefore, this results in
\begin{equation}
\langle \psi_1^{\dagger}(\tau)\psi_1(0)\rangle + \langle \psi_1(\tau)\psi_1^{\dagger}(0)\rangle = \frac{1}{M}\sum_{k=[-\pi;\pi]} e^{-\frac{E(k)\tau}{\hbar}}.
\end{equation}
The quantum phase transition implies a zero of energy $E(k)$ at $k=\pi$. If we would keep only this mode in the sum, then the prefactor is small $\sim \frac{1}{M}$. Therefore, we will develop the integral close to $k=\pi$ within the continuum limit 
such that for $J_1=J_2=J$, we have
\begin{equation}
E(k) = 2J\left | \cos\frac{k}{2} \right|.
\end{equation}
In this way,
\begin{equation}
\langle \psi_1^{\dagger}(\tau)\psi_1(0)\rangle + \langle \psi_1(\tau)\psi_1^{\dagger}(0)\rangle = \frac{1}{\pi}\int_0^{\pi} e^{-\frac{2J}{\hbar}\cos \frac{k}{2}\tau} dk.
\end{equation}
If we do the change of variables $u=\frac{2J}{\hbar}\tau \cos\frac{k}{2}$, this results in the integral
\begin{equation}
\langle \psi_1^{\dagger}(\tau)\psi_1(0)\rangle + \langle \psi_1(\tau)\psi_1^{\dagger}(0)\rangle = -\frac{1}{\pi} \frac{\hbar}{J\tau} \int \frac{e^{-u}}{\sqrt{1-\frac{\hbar^2 u^2}{4 J^2 \tau^2}}} du.
\end{equation}
At this stage, the evaluation is exact. 

Now, we need to clarify the interval of validity of the integral variable. When $k=\pi$ then $u=0$. The dominant response should be close to $k=\pi$. When $k=0$, the primitive gives a less dominant response. Therefore, if we develop the function close to $\pi$, say within an interval
$k\in[\pi-\alpha;\pi]$ then this corresponds to $u\in [\frac{J\tau}{\hbar} \alpha;0]$. If we develop the function for the long time limit that will dominate the response at low temperature, we obtain
\begin{equation}
\langle \psi_1^{\dagger}(\tau)\psi_1(0)\rangle + \langle \psi_1(\tau)\psi_1^{\dagger}(0)\rangle \approx -\frac{1}{\pi} \frac{\hbar}{J\tau} \int_{\frac{J\tau \alpha}{\hbar}}^0 e^{-u} du,
\end{equation}
and
\begin{equation}
\langle \psi_1^{\dagger}(\tau)\psi_1(0)\rangle + \langle \psi_1(\tau)\psi_1^{\dagger}(0)\rangle \approx \frac{1}{\pi} \frac{\hbar}{J\tau}+{\cal O}(e^{-\frac{\hbar J \alpha}{\hbar}}).
\end{equation}

Therefore, at long times, we verify that the dominant term comes from the limit $k=\pi$ or $u=0$ for this integral. Then, when the magnetic field $h_1\rightarrow 0$ we obtain
\begin{equation}
\chi_1 \approx +\frac{1}{\pi} \frac{\hbar}{J}\int_{\frac{\hbar}{J}}^{\frac{\hbar}{|J_1-J_2|}} \frac{d\tau}{\tau}.
\end{equation}
Therefore, when $h_1\rightarrow 0$ and we are very close to the quantum phase transition, 
\begin{equation}
\chi_1 \approx -\frac{1}{\pi} \frac{\hbar}{J}\ln |J_1-J_2|.
\end{equation}
This gives rise to a linear edge magnetization with an infinite logarithmic singularity close to the critical point, e.g. for $J_1<J_2$
\begin{equation}
\langle \sigma_1^z\rangle = -\frac{1}{\pi}\frac{\hbar}{J}\ln(J_2-J_1) h_1.
\end{equation}
It should be noted that in the case where both $c_1$ and $d_1$ would interact strongly with the conduction band, i.e. they would show the same Green's function then the free energy correction would be
\begin{equation}
\Delta F_1 \propto h_1^2 \frac{\hbar^2}{J^2} \int_{\frac{\hbar}{J}}^{\beta} \frac{d\tau}{\tau^2} \rightarrow h_1^2\frac{\hbar^2}{J^2}\left(-\frac{1}{\beta} +\frac{J}{\hbar}\right).
\end{equation}
In the limit where $\beta\rightarrow +\infty$ the local spin susceptibility tends to a constant term $\chi_1\sim \frac{\hbar}{J}$.

\begin{figure}[t]
\includegraphics[width=8cm]{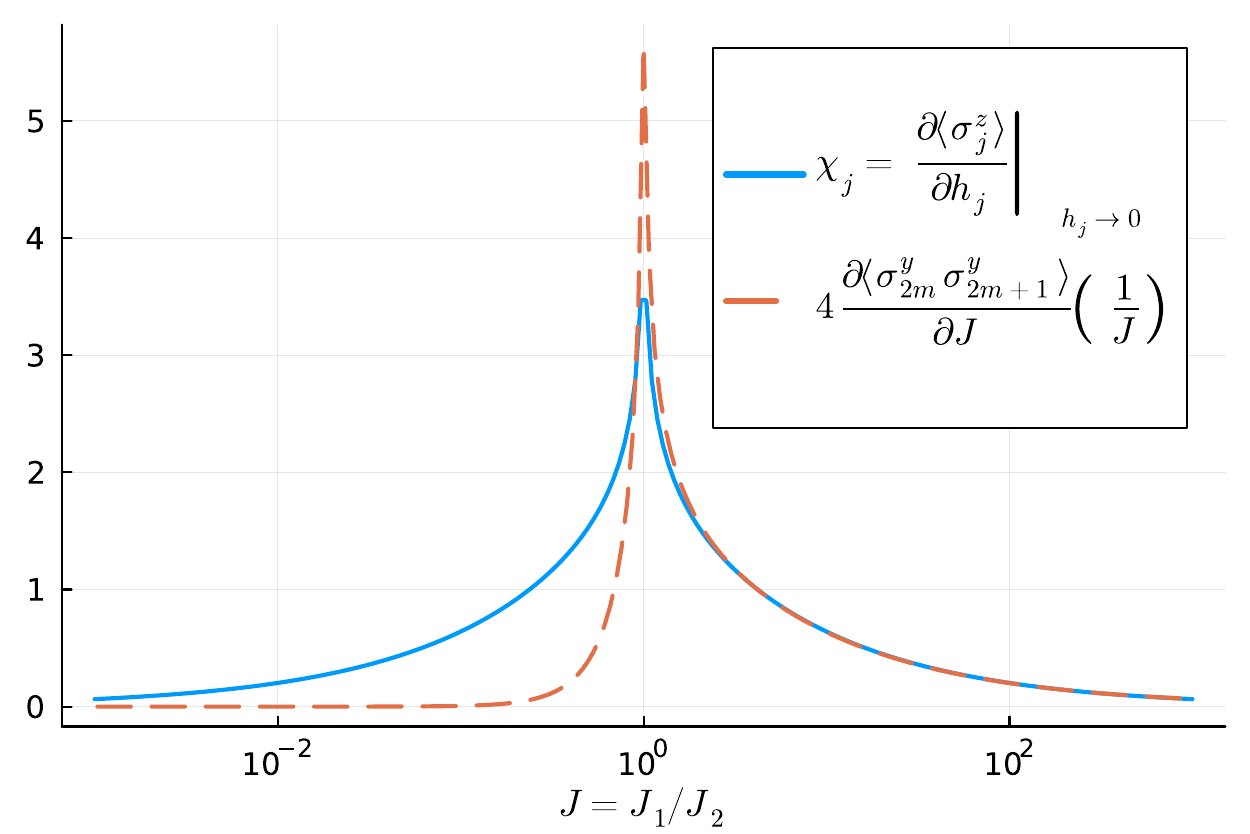}
\includegraphics[width=8.5cm]{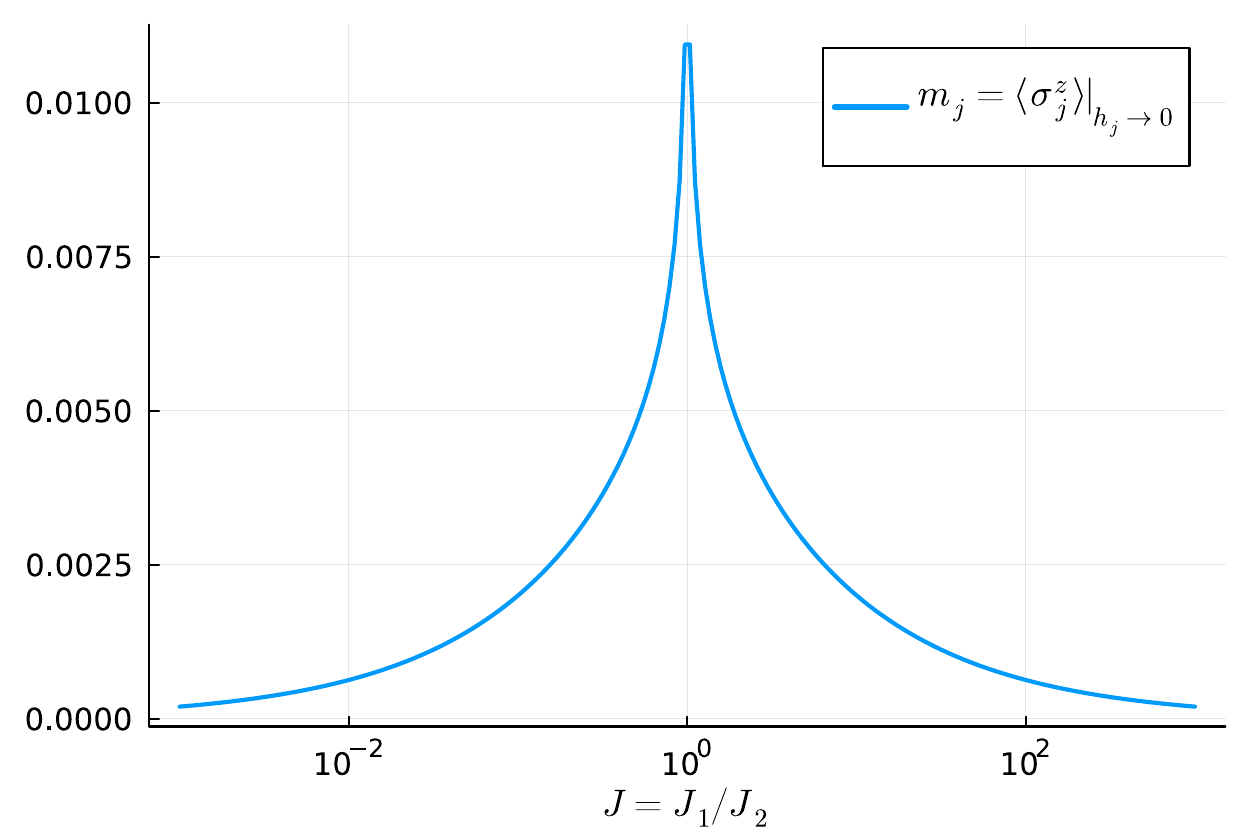}
\caption{Local spin susceptibility in the bulk with DMRG (400 sites) and bulk magnetization when the local magnetic field is very small.}
\label{Figure}
\end{figure}

We mention here that an evaluation of the local susceptibility can be done at any site with a similar result at the transition. The bulk magnetization along $z$ direction goes to zero when $J_1/J_2\rightarrow 0$ and when $J_2/J_1\rightarrow +\infty$. We present the DMRG results in Fig. \ref{Figure} in the bulk for the local susceptibility and bulk magnetization. 

\vskip 0.5cm

\hskip -0.3cm [1] J. Hutchinson and K. Le Hur, Communications Physics {\bf 4}, 144 (2021).
\\
\hskip -0.3cm [2]  P. Roushan, C. Neill, Y. Chen {\it et al.} Nature {\bf 515} 241-244 (2014); M. D. Schroer, M. H. Kolodrubetz, W. F. Kindel {\it et al.} Phys. Rev. Lett. {\bf 113} 050402 (2014).
\\
\hskip -0.3cm [3] K. Le Hur, Phys. Rev. B {\bf 108}, 235144 (2023); 
K. Le Hur, Physics Reports  {\bf 1104}, 1-42 (2025).
\\
\hskip -0.3cm [4] K. Le Hur, arXiv:2209.15381, 108 pages.

\end{document}